# APPROXIMATE SOLUTIONS OF THE FRACTIONAL SCHRODINGER EQUATION FOR THE SCREENED KRATZER POTENTIAL

U. S. Okorie[1&3*], A. N. Ikot[2&1], P. O. Amadi[1], G. J. Rampho[2]

[1]Theoretical Physics Group, Department of Physics, University of Port Harcourt, P. M. B. 5323 Choba, Nigeria

[2]Department of Physics, University of South Africa, Florida 1710, Johannesburg, South Africa

[3]Department of Physics, Akwa Ibom State University, Ikot Akpaden P. M. B. 1167, Uyo, Nigeria

**Abstract:**

By employing the concept of conformable fractional Nikiforv-Uvarov (NU) method, we solved the fractional Schrodinger equation with the screened Kratzer potential (SKP). By applying the Greene-Aldrich approximation and a coordinate transformation schemes, the analytical expressions of the bound state energy spectra and eigenfunctions for SKP were obtained. Numerical results for the energies of SKP for lithium hydride and hydrogen chloride diatomic molecules were computed for different fractional parameters. Also, the graphical variation of the bound state energy eigenvalues of SKP for LiH with different potential parameters and quantum numbers were discussed, as regards the selected fractional parameters. Our results are new and have not been reported in any literature before.

Corresponding Author's Email: uduakobongokorie@aksu.edu.ng

1. **Introduction**

One of the molecular potential functions widely studied by numerous researchers is the Kratzer potential [1]. The Kratzer potential is known to have a long-range attraction and a repulsive part. The integration of these parts makes the potential reliable, as far as its vibrational and rotational energy eigenvalues are concerned [2]. The Kratzer potential also approaches infinity as the inter-nuclear distance approaches zero. This is because of the repulsion that exists between the molecules of the potential [3]. This potential has been used to describe molecular structure between two atoms [4-9]. Its applications span into quantum chemistry and atomic and molecular physics [10]. By combining an exponential term to the standard Kratzer potential, we earlier proposed the screened Kratzer potential of the form [11]

$$V(r) = -2D_e \left( \frac{A}{r} - \frac{B}{2r^2} \right) e^{-\delta r}, \ A \equiv r_e, \ B \equiv r_e^2 \qquad (1),$$

with $D_e$, $r_e$ and $\delta$ being the dissociation energy, equilibrium bond length and screening parameter. This was done to obtain a wider range of applications [12].

Different mathematical approaches have been employed to obtain the energy eigensolutions for Kratzer-like potentials and other molecular potential systems. These include suppersymmetric quantum mechanics (SUSYQM) approach [13-16], exact and proper quantization [17-20], asymptotic iteration method (AIM) [21, 22], factorization method [23-26], Nikiforov-Uvarov (NU) method [27-30] and others.

The concept of fractional calculus is based on non-integer order differentiation and integration [31]. Before now, fractional calculus has attracted attention of many authors due to its applications in various fields of science and engineering [32]. In physics and its related areas, for instance, Laskin [33] studied the fractional Schrodinger equation containing Caputo fractional derivative and the quantum Riesz fractional operator. Taskin [34] considered the fractional derivative with Schrodinger equation in the non-relativistic quantum mechanics. Of recent, Karayer *et al.* [35] derived the conformable fractional form of the Nikiforov-Uvarov (NU) method and used it to solve the local fractional Schrodinger equation for harmonic oscillator potential, Hulthen potential and Woods-Saxon potential. Also, using the fractional form of NU method, Karayer *et al.* [36] investigated the analytical solutions of local Klein-Gordon equation for the generalized Hulthen potential. Baleanu *et al.* [37] also presented the applications of fractional calculus in areas of complex and nonlinear physics. In another development, Al-Jamel [38] studied the heavy quarkonium energy spectra within the framework of fractional Schrodinger equation with extended Cornell potential model. Conformable fractional derivative and integral have been used to study the fractional version of the Newtonian mechanics [39]. In the fractional calculus and its application in quantum mechanics, the fractional parameter $0 < \alpha \leq 1$ is related to the roughness character of the space-time. Also, the nature of solutions of the wave equations at various values of the fractional parameter $\alpha$ signifies the underlying behaviour of the quantum mechanical systems [40]. For $\alpha = 1$, the quantum mechanical system reduces to the normal quantum mechanics.

We are motivated to consider the approximate solutions of the fractional Schrodinger equation for the screened Kratzer potential. This work is novel and has been reported anywhere, to the best of our knowledge.

The paper is organized as follows. In section 2, we present the conventional Schrodinger equation formalism with the screened Kratzer potential. In Section 3, the theory of conformable fractional Nikiforov-Uvarov method is presented briefly. The eigensolutions of the conformable fractional Schrodinger equation with the screened Kratzer potential is presented in section 4. Discussion of results is given in section 5. Finally, the conclusion remarks are presented in section 6.

## 2. Schrodinger Equation Formalism with the Screened Kratzer Potential

The radial Schrodinger equation is often defined as [41]:

$$\frac{d^2 R_{n\ell}(r)}{dr^2} + \frac{2\mu}{\hbar^2}\left[E_{n\ell} - V(r) - \frac{\ell(\ell+1)\hbar^2}{2\mu r^2}\right]R_{n\ell}(r) = 0 \qquad (2),$$

where $\mu$ is the reduced mass, $E_{n\ell}$ is the energy eigenvalues of the screened Kratzer potential to be determined, $\hbar$ is the reduced Planck's constant and $n$ and $\ell$ are the radial and orbital angular momentum quantum numbers respectively (also known as the vibration-rotation quantum numbers) [14]. Substituting Eq. (1) into Eq. (2) results:

$$\frac{d^2 R_{n\ell}(r)}{dr^2} + \left[\frac{2\mu E_{n\ell}}{\hbar^2} + \frac{2\mu D_e}{\hbar^2}\left(\frac{A}{r} - \frac{B}{2r^2}\right)e^{-\delta r} - \frac{\ell(\ell+1)}{r^2}\right]R_{n\ell}(r) = 0 \qquad (3).$$

Eq. (3) cannot be solved analytically for $\ell \neq 0$ due to the presence of the centrifugal term. Therefore, we employ the Greene-Aldrich approximation to the centrifugal term as given below [42]

$$\frac{1}{r^2} \approx \frac{\delta^2}{\left(1-e^{-\delta r}\right)^2}, \quad \frac{1}{r} \approx \frac{\delta}{\left(1-e^{-\delta r}\right)} \qquad (4).$$

Substituting the approximation scheme given above and employing a coordinate transformation of the form $z = e^{-\delta r}$ in Eq. (3), the differential equation of the form is obtained:

$$\frac{d^2 R_{n\ell}(z)}{dz^2} + \frac{(1-z)}{z(1-z)}\frac{dR_{n\ell}(z)}{dz} + \frac{1}{z^2(1-z)^2}\left[-\left(\varepsilon_{n\ell}^2+\gamma\right)z^2 + \left(2\varepsilon_{n\ell}^2+\gamma-K\right)z - \left(\varepsilon_{n\ell}^2+\eta\right)\right]R_{n\ell}(z) = 0$$
(5).

Here, we have adopted the following definition of terms:

$$-\varepsilon_{n\ell}^2 = \frac{2\mu E_{n\ell}}{\hbar^2 \delta^2};\ \gamma = \frac{4\mu D_e A}{\hbar^2 \delta};\ K = \frac{2\mu D_e B}{\hbar^2};\ \eta = \ell(\ell+1) \qquad (6).$$

### 3. Theory of Conformable Fractional Nikiforov Uvarov (NU) Method

The conformable fractional Nikiforov-Uvarov (NU) method was introduced to solve the conformable fractional order Schrodinger equation [43]. This method uses the key property of the conformable fractional derivative operator given as [44]

$$D^{\varphi}\left[f(x)\right] = \lim_{\varsigma \to 0}\frac{f\left(x+\sigma x^{1-\varphi}\right) - f(x)}{\varsigma}, \quad x > 0 \qquad (7),$$

$$f^{(\varphi)}(0) = \lim_{\varsigma \to 0} f^{(\varphi)}(x) \qquad (8).$$

Here, $0 < \varphi \leq 1$ and $D^{\varphi}$ is known as the local fractional derivative operator [44]. The basic rules of the local fractional derivative operator which are valid in standard calculus are defined as

$$D^{\varphi}[ag + bh] = aD^{\varphi}[g] + bD^{\varphi}[h] \qquad \textbf{\textit{(Linearity)}} \qquad (9a),$$

$$D^{\varphi}[gh] = gD^{\varphi}[h] + hD^{\varphi}[g] \qquad \textbf{\textit{(Product Rule)}} \qquad (9b),$$

$$D^{\varphi}[g(h)] = \frac{dg}{dh}D^{\varphi}[h] \qquad \textbf{\textit{(Chain Rule)}} \qquad (9c),$$

$$D^{\varphi}[g] = x^{1-\varphi}\frac{dg}{dx} \qquad \textbf{\textit{(Key Property)}} \qquad (9d).$$

If $g$ is differentiable, then the $\varphi th$ order derivative of $g$ would be equal to the product of its first-order derivative with $x^{1-\varphi}$.

By employing the key property of the conformable fractional derivative operator, the following second-order differential equation via NU method [27] is obtained:

$$\frac{d^2\psi(z)}{dz^2} + \frac{\tilde{\tau}_f(z)}{\sigma_f(z)}\frac{d\psi(z)}{dz} + \frac{\tilde{\sigma}_f(z)}{\sigma_f^2(z)}\psi(z) = 0 \qquad (10),$$

where $\tilde{\tau}_f(z) = (1-\varphi)z^{-\varphi}\sigma(z) + \tilde{\tau}(z)$, $\sigma_f(z) = z^{1-\varphi}\sigma(z)$ and the subscript $f$ represents the term 'fractional'. Also, $\tilde{\tau}_f(z)$ is a function of at most $\varphi th$ degree, $\sigma_f(z)$ is a function of at most $(\varphi+1)th$ degree and $\tilde{\sigma}(z)$ is a function of at most $2\varphi th$ degree.

With the help of the newly defined functions related to the initial functions in the basic NU equation, the eigensolutions of Eq. (10) can be obtained thus:

$$\pi_f(z) = \frac{\sigma'_f(z) - \tilde{\tau}_f(z)}{2} \pm \sqrt{\left(\frac{\sigma'_f(z) - \tilde{\tau}_f(z)}{2}\right)^2 - \tilde{\sigma}(z) + \left(k(z)\sigma_f(z)\right)} \qquad (11).$$

The expression under the square root sign of Eq. (11) must be square of a $\varphi th$- order function. Here, choosing the function $k(z)$ properly, we obtain the following expressions:

$$\tau_f(z) = \tilde{\tau}_f(z) + 2\pi_f(z) \qquad (12),$$

$$\lambda(z) = k(z) + \pi'_f(z) \qquad (13),$$

$$\lambda_n(z) = -n\tilde{\tau}'_f(z) - \frac{n(n-1)}{2}\sigma''_f(z), \ n = 0, 1, 2, ... \qquad (14).$$

The eigenvalue solution can be obtained when Eq. (13) is equal to Eq. (14). To obtain the eigenfunction solution, we define the following:

$$\frac{\phi'(z)}{\phi(z)} = \frac{\pi_f(z)}{\sigma_f(z)} \qquad (15),$$

$$\left(\sigma_f(z)\rho(z)\right)' = \tau_f(z)\rho(z) \qquad (16),$$

$$y_{n\ell}(z) = \frac{B_n}{\rho(z)}\frac{d^n}{dz^n}\left[\sigma_f^n(z)\rho(z)\right] \qquad (17),$$

where $\psi(z) = \phi(z)y(z)$. The complete explanation of the conformable fractional NU method can be obtained in Ref. [35].

## 4. Eigensolutions of the Conformable Fractional SE with the Screened Kratzer Potential

By replacing integer orders by fractional orders in Eq. (5), the conformable fractional form of the Schrodinger equation for the screened Kratzer potential is written as

$$D^{\varphi}\left[D^{\varphi}R_{n\ell}(z)\right] + \frac{(1-z^{\varphi})}{z^{\varphi}(1-z^{\varphi})}D^{\varphi}\left[R_{n\ell}(z)\right] + \frac{1}{z^{2\varphi}(1-z^{\varphi})^2}\left[-\zeta_1 z^{2\varphi} + \zeta_2 z^{\varphi} - \zeta_3\right]R_{n\ell}(z) \quad (18),$$

where

$$\zeta_1 = \varepsilon_{n\ell}^2 + \gamma; \quad \zeta_2 = 2\varepsilon_{n\ell}^2 + \gamma - K; \quad \zeta_3 = \varepsilon_{n\ell}^2 + \eta \quad (19).$$

By employing the key property of Eq. (9d), Eq. (18) is transformed into a second-order differential equation of the form:

$$\frac{d^2 R_{n\ell}(z)}{dz^2} + \frac{(2-\varphi)(1-z^{\varphi})}{z(1-z^{\varphi})}\frac{dR_{n\ell}(z)}{dz} + \frac{1}{z^2(1-z^{\varphi})^2}\left[-\zeta_1 z^{2\varphi} + \zeta_2 z^{\varphi} - \zeta_3\right]R_{n\ell}(z) = 0 \quad (20).$$

Comparing Eq. (20) with Eq. (10), we obtain the following expressions:

$$\tilde{\tau}_f(z) = (2-\varphi)(1-z^{\varphi}) \quad (21),$$

$$\sigma_f(z) = z(1-z^{\varphi}) \quad (22),$$

$$\tilde{\sigma}_f(z) = -\zeta_1 z^{2\varphi} + \zeta_2 z^{\varphi} - \zeta_3 \quad (23).$$

Substituting Eqs. (21) – (23) into Eq. (11) give

$$\pi_f(z) = \frac{(\varphi-1)+(1-2\varphi)z^{\varphi}}{2} \pm \frac{1}{2}\sqrt{\left[\frac{(1-2\varphi)^2}{4}+\frac{(\varphi-1)^2}{4}+\zeta_1 - k\, z^{1-\varphi}\right]z^{2\varphi} + \left[\frac{(\varphi-1)(1-2\varphi)}{2}-\zeta_2 + k\, z^{1-\varphi}\right]z^{\varphi} + \zeta_3} \quad (24).$$

To find the constant $k$, the discriminant of the expression under the square root of Eq. (24) must be zero. Hence,

$$k_{\pm} = \left[-(Q+2R) \pm 2\sqrt{R(R+Q+P)}\right]z^{\varphi-1} \quad (25).$$

Here, $P$, $Q$ and $R$ are defined as

$$P = \frac{(1-2\varphi)^2}{4} + \frac{(\varphi-1)^2}{4} + \zeta_1; \quad Q = \frac{(\varphi-1)(1-2\varphi)}{2} - \zeta_2; \quad R = \zeta_3 \quad (26).$$

By employing Eq. (26), Eq. (24) becomes

$$\pi_f(z) = \frac{(\varphi-1)+(1-2\varphi)z^\varphi}{2} \pm \begin{cases} \left(\sqrt{R}-\sqrt{R+Q+P}\right)z^\varphi - \sqrt{R}, & k_+ = \left[-(Q+2R)+2\sqrt{R(R+Q+P)}\right]z^{\varphi-1} \\ \left(\sqrt{R}+\sqrt{R+Q+P}\right)z^\varphi - \sqrt{R}, & k_- = \left[-(Q+2R)-2\sqrt{R(R+Q+P)}\right]z^{\varphi-1} \end{cases}$$

(27).

By choosing $k_-$ the function $\pi_f(z)$ becomes

$$\pi_f(z) = \frac{(\varphi-1)+(1-2\varphi)z^\varphi}{2} - \left\{\left(\sqrt{R}+\sqrt{R+Q+P}\right)z^\varphi - \sqrt{R}\right\} \quad (28).$$

Also, the expressions for $\tau_f(z)$, $\lambda(z)$ and $\lambda_n(z)$ are obtained respectively as

$$\tau_f(z) = \left[2\sqrt{R}+(\varphi-1)+(2-\varphi)\right] + \left[1-2\varphi-2\left(\sqrt{R}+\sqrt{R+Q+P}\right)-(2-\varphi)\right]z^\varphi \quad (29),$$

$$\lambda(z) = \left[-Q-2R-2\sqrt{R(R+Q+P)}+\frac{\varphi(1-2\varphi)}{2}-\varphi\sqrt{R}-\varphi\sqrt{R+Q+P}\right]z^{\varphi-1} \quad (30),$$

$$\lambda_n(z) = \varphi n\left[1+\varphi+2\left(\sqrt{R}+\sqrt{R+Q+P}\right)\right]z^{\varphi-1} + \frac{n(n-1)\varphi(\varphi+1)}{2}z^{\varphi-1} \quad (31).$$

Equating Eq. (30) and Eq. (31) and using Eqs. (6), (19) and (26), we obtain the energy eigenvalue expression as

$$E_{n\ell} = \frac{\ell(\ell+1)\hbar^2\delta^2}{2\mu} - \frac{\hbar^2\delta^2}{2\mu}\left[\frac{\frac{\varphi(\varphi+1)}{2}\left(n+\frac{1}{2}+\frac{2\sqrt{\omega}}{(\varphi+1)}\right)^2 - \left(\frac{\varphi(1-2\varphi)}{2}-\Lambda+\frac{\varphi(\varphi+1)}{8}+\frac{2\varphi\omega}{(\varphi+1)}\right)^2}{2\varphi\left(n+\frac{1}{2}+\frac{\sqrt{\omega}}{\varphi}\right)}\right]^2$$

(32),

where

$$\omega = \frac{(1-2\varphi)^2}{4} + \frac{(\varphi-1)^2}{4} + \frac{(\varphi-1)(1-2\varphi)}{2} + \frac{2\mu D_e B}{\hbar^2} + \ell(\ell+1) \quad (33)$$

$$\Lambda = \frac{(\varphi-1)(1-2\varphi)}{2} + \left(\frac{2\mu D_e B}{\hbar^2} - \frac{4\mu D_e A}{\hbar^2\delta}\right) + 2\ell(\ell+1) \quad (34).$$

When the fractional parameter $\varphi = 1$, the result of the energy eigenvalue obtained is equivalent with that in Ref. [11] for screened Kratzer potential using the standard NU method.

To obtain the eigenfunction solution, firstly, Eq. (15) is employed to determine $\phi(z)$ as

$$\phi(z) = z^{\left(\frac{\varphi-1}{2}\right)} \left(1 - z^{\varphi}\right)^{\left(\frac{\sqrt{\omega}}{\varphi} + \frac{1}{2}\right)} \tag{35}$$

By using Eqs. (16) and (17), the expressions for $\rho(z)$ and $y_{n\ell}(z)$ are obtained respectively as

$$\rho(z) = z^{2\sqrt{R}} \left(1 - z^{\varphi}\right)^{\left(\frac{\sqrt{\omega}}{\varphi} + 2\left(\frac{1-2\varphi}{\varphi}\right)\right)} \tag{36},$$

$$y_{n\ell}(z) = \mathbb{N}_{n\ell}\, z^{-2\sqrt{R}} \left(1 - z^{\varphi}\right)^{-\left(\frac{\sqrt{\omega}}{\varphi} + 2\left(\frac{1-2\varphi}{\varphi}\right)\right)} \frac{d^n}{dz^n}\left[ z^{n+2\sqrt{R}} \left(1 - z^{\varphi}\right)^{n + \left(\frac{\sqrt{\omega}}{\varphi} + 2\left(\frac{1-2\varphi}{\varphi}\right)\right)} \right] \tag{37}.$$

Using the transformation $\psi_{n\ell}(z) = \phi(z)\, y_{n\ell}(z)$, we obtain the eigenfunction solution as

$$\psi_{n\ell}(z) = \mathbb{N}_{n\ell}\, z^{\left(\frac{\varphi-1}{2} - 2\sqrt{R}\right)} \left(1 - z^{\varphi}\right)^{\left(\frac{1}{2} - 2\left(\frac{1-2\varphi}{\varphi}\right)\right)} \frac{d^n}{dz^n}\left[ z^{(n+2\sqrt{R})} \left(1 - z^{\varphi}\right)^{\left(n + \frac{\sqrt{\omega}}{\varphi} + 2\left(\frac{1-2\varphi}{\varphi}\right)\right)} \right] \tag{38}.$$

## 5. Results and Discussion

With the help of Eq. (32), we present the numerical results of the rotation-vibrational energy eigenvalues for lithium hydride (LiH) and hydrogen chloride (HCl) diatomic molecules in this work. The spectroscopic parameters for the selected diatomic molecules are given in Table 1. Also, the following conversions were useful in our computations [45]:

$1\, amu = 931.494028 \frac{MeV}{c^2}$ ; $\hbar c = 1973.29\, eV\, \text{Å}$. We have chosen the following fractional parameters arbitrarily for discussion purposes: $\alpha = 0.2, 0.5, 0.7$ and $1$. Tables 2 and 3 show that for LiH and HCl respectively, the bound state energy eigenvalues increase as the rotational quantum number increases for any vibrational quantum number. Also, the bound state energy eigenvalues for the selected diatomic molecules decrease with increase in vibrational quantum numbers considered. The behaviour of the bound state energy eigenvalues with the rotational and vibrational quantum numbers are seen to be the same for the fractional parameters considered. In addition, the bound state energy eigenvalues for the selected diatomic molecules at different quantum states decrease as the fractional parameter inceases.

The variation of the energy eigenvalues of screened kratzer potential (SKP) with different potential parameters and quantum numbers were analyzed graphically for lithium hydride diatomic molecules at various fractional parameters. Figure 1 shows a gradual decrease in the energy eigenvalues of SKP as the dissociation energy is increased. Also, the energy eigenvalues of SKP is decreased as the fractional parameter is increased. Figure 2 shows the variation of the energy eigenvalues of SKP with equilibrium bond length for various fractional parameters. Here, the energy eigenvalues decrease first to a unique value for each fractional parameter as the equilibrium bond length is $0 < r_e \leq 0.2\, \text{Å}$. As $r_e$ increases beyond $0.2 \overset{o}{\text{A}}$ to $1.0 \overset{o}{\text{A}}$, the bound state energy eigenvalues increase to zero. As $r_e$ is further enhanced beyond $1.0 \overset{o}{\text{A}}$, the bound state energy eigenvalues decrease gradually. In Figure 3, we observe a gradual increase in the bound state energy eigenvalues of SKP as the screening parameter increases for the selected fractional parameters. As the screening parameter is enhanced beyond $1.0 \overset{o}{\text{A}}^{-1}$, the bound state energy eigenvalues begin to decrease gradually. Figure 4 shows the variation of the bound state energy eigenvalues of SKP with the reduced mass. Here, the bound state energy first increased sharply at zero reduced mass and later increased monotonously as the reduced mass is further enhanced. In Figure 5, the bound state energy eigenvalues reduce as the vibrational quantum number is increased for the fractional parameters selected. Figure 6 shows a gradual increase in the bound state energy eigenvalues of SKP as the rotational quantum number is increased.

## 6. Concluding Remarks

In this research, we solved the radial fractional Schrodinger equation for the screened Kratzer potential and obtained analytical expressions for bound state energy eigenvalues and eigenfunctions using the conformable fractional Nikforov-Uvarov (NU) method. We also presented the numerical results of the bound state energies for lithium hydride and hydrogen chloride diatomic molecules at different fractional parameters, by employing their spectroscopic parameters. In addition, the variation of the bound state energy eigenvalues of screened Kratzer potential for lithium hydride with dissociation energy, equilibrium bond length, screening parameter, reduced mass, vibrational and rotational quantum numbers have been graphically discussed at different fractional parameters chosen. Due to the fact that our study is novel, we cannot compare our results due to lack of related studies in available literatures.

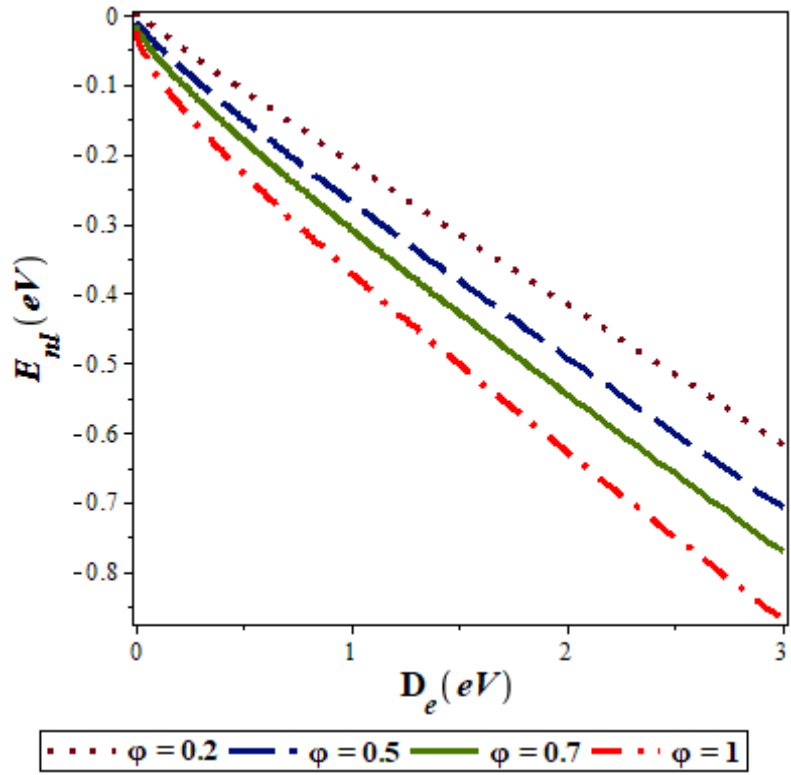

Figure 1: Energy eigenvalues of SKP versus dissociation energy for LiH at various $\varphi$.

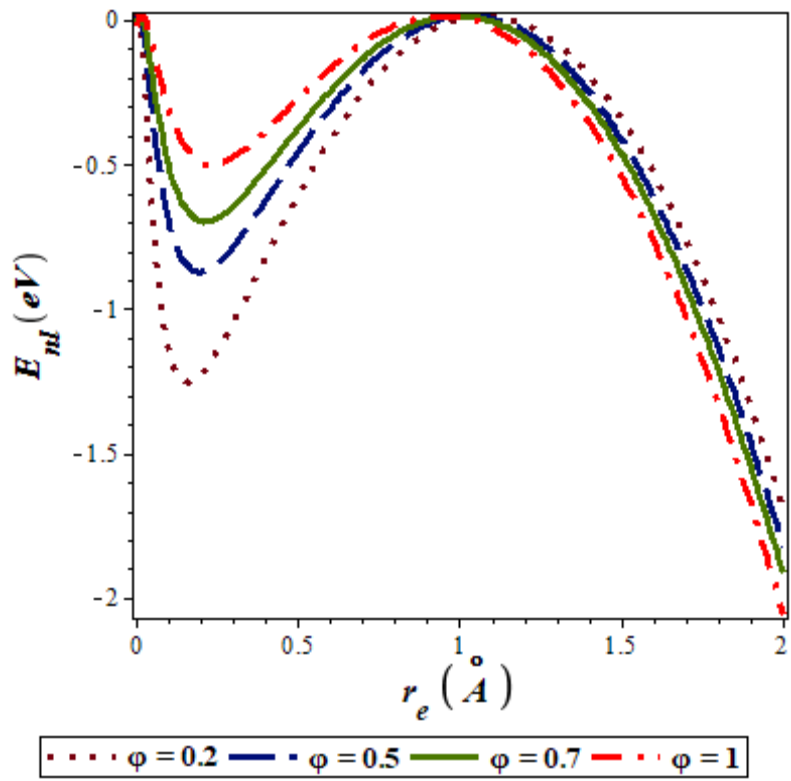

Figure 2: Energy eigenvalues of SKP versus equilibrium bond length for LiH at various $\varphi$.

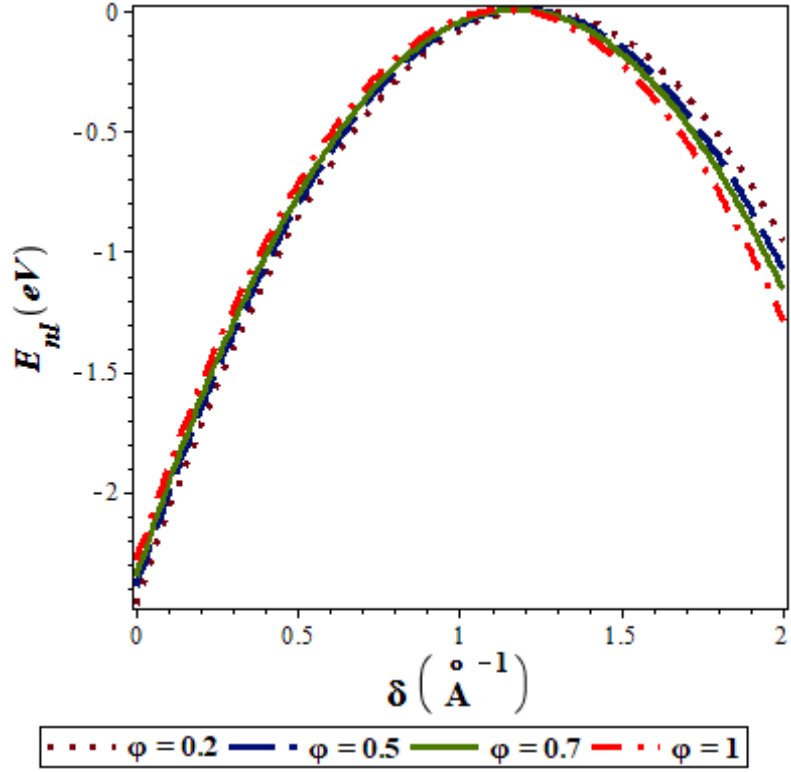

Figure 3: Energy eigenvalues of SKP versus screening parameter for LiH at various $\varphi$.

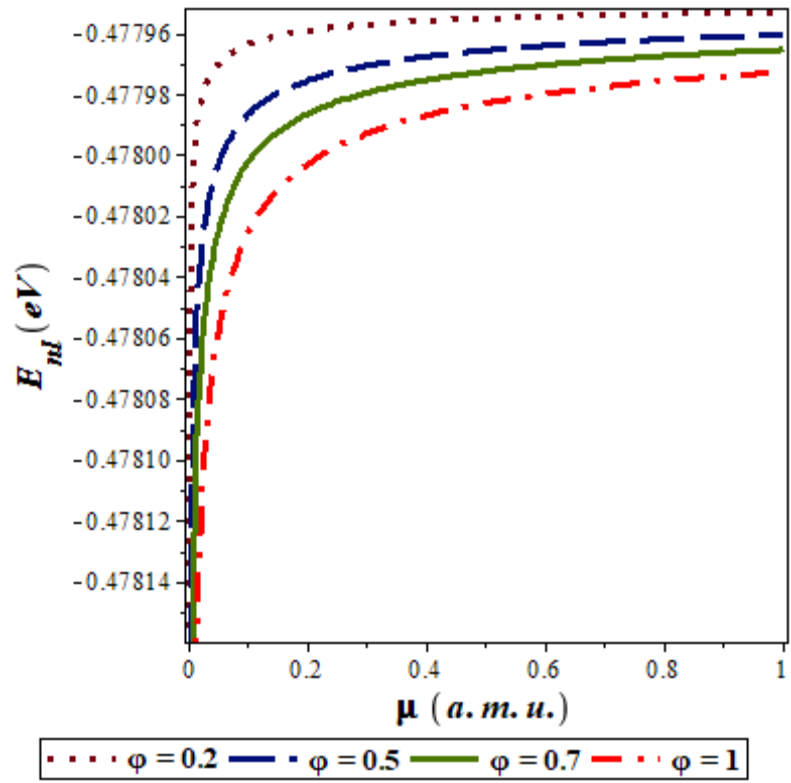

Figure 4: Energy eigenvalues of SKP versus reduced mass for LiH at various $\varphi$.

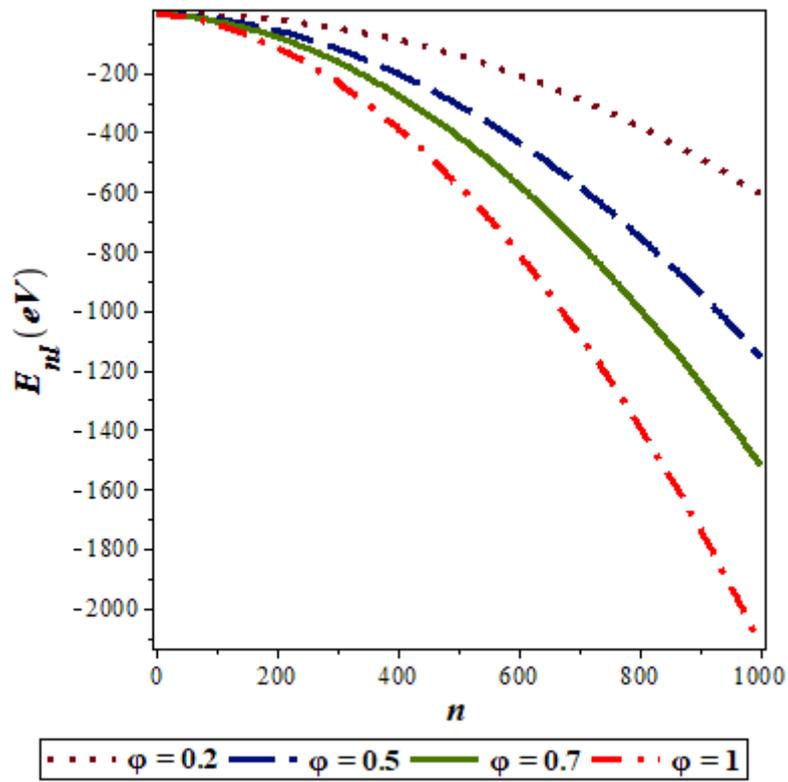

Figure 5: Energy eigenvalues of SKP versus vibrational quantum number for LiH at various $\varphi$.

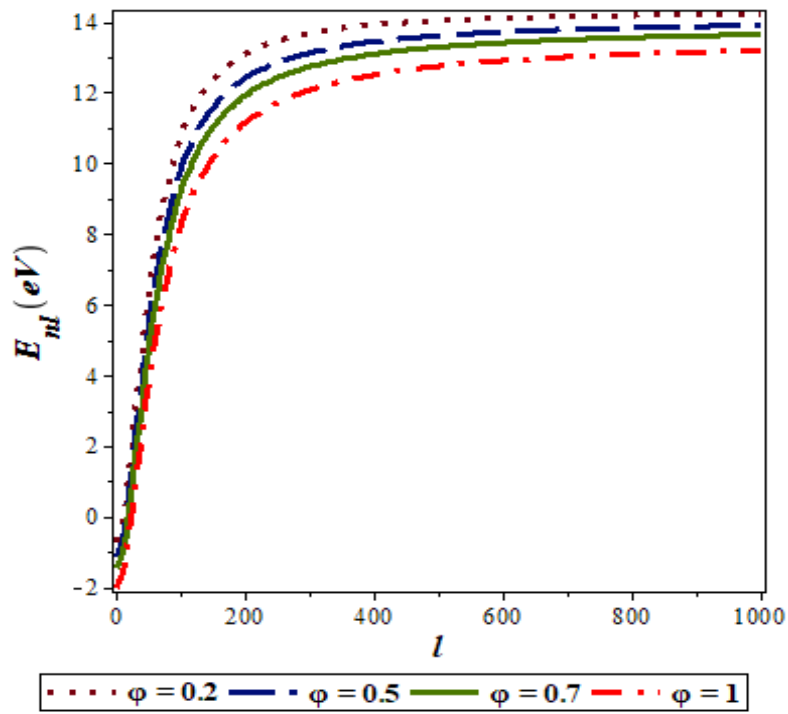

Figure 6: Energy eigenvalues of SKP versus rotational quantum number for LiH at various $\varphi$

**Table 1:** Spectroscopic Parameter for the selected diatomic molecules [45]

| Molecules | $D_e(eV)$ | $r_e(\text{Å})$ | $\alpha(\text{Å}^{-1})$ | $\mu(a.m.u.)$ |
|---|---|---|---|---|
| $HC\ell$ | 4.61907 | 1.2746 | 2.38057 | 0.9801045 |
| $LiH$ | 2.515287 | 1.5956 | 1.7998368 | 0.8801221 |

**Table 2**: Energy Eigenvalues $\left(-E_{nl}(eV)\right)$ of the screened Kratzer potential for *LiH* diatomic molecule with different values of $\varphi$.

| n | l | $\varphi = 0.2$ | $\varphi = 0.5$ | $\varphi = 0.7$ | $\varphi = 1$ |
|---|---|---|---|---|---|
| 0 | 0 | 0.4879150312 | 0.5038728265 | 0.5145976750 | 0.5308115940 |
| 1 | 0 | 0.5090494540 | 0.5579453710 | 0.5914247655 | 0.6429174030 |
|   | 1 | 0.4981027183 | 0.5471699748 | 0.5807608048 | 0.6324165643 |
| 2 | 0 | 0.5307580615 | 0.6146487010 | 0.6729781445 | 0.7638658880 |
|   | 1 | 0.5198888058 | 0.6040597668 | 0.6625687948 | 0.7537156258 |
|   | 2 | 0.4981980770 | 0.5829277640 | 0.6417947450 | 0.7334580135 |
| 3 | 0 | 0.5530472055 | 0.6739687995 | 0.7591674000 | 0.8932997865 |
|   | 1 | 0.5422553768 | 0.6635627988 | 0.7490049038 | 0.8834836783 |
|   | 2 | 0.5207189735 | 0.6427954290 | 0.7287229195 | 0.8638921855 |
|   | 3 | 0.4885319856 | 0.6117554826 | 0.6984070006 | 0.8346063226 |
| 4 | 0 | 0.5759231830 | 0.7358922080 | 0.8499074740 | 1.030893206 |
|   | 1 | 0.5652087208 | 0.7256657203 | 0.8399844753 | 1.021396095 |
|   | 2 | 0.5438265275 | 0.7052561895 | 0.8201799185 | 1.002440549 |
|   | 3 | 0.5118695516 | 0.6747500370 | 0.7905762386 | 0.9741035150 |
|   | 4 | 0.4694759514 | 0.6342757439 | 0.7512960069 | 0.9364994244 |
| 5 | 0 | 0.5993922390 | 0.8004059865 | 0.9451182945 | 1.176348380 |
|   | 1 | 0.5887550778 | 0.7903557063 | 0.9354278098 | 1.167156252 |
|   | 2 | 0.5675269640 | 0.7702974330 | 0.9160867795 | 1.148808761 |
|   | 3 | 0.5357998176 | 0.7403153010 | 0.8871746660 | 1.121379040 |
|   | 4 | 0.4937102649 | 0.7005343834 | 0.8488096244 | 1.084975877 |
|   | 5 | 0.4414381716 | 0.6511193911 | 0.8011472786 | 1.039742577 |

**Table 3**: Energy Eigenvalues $\left(-E_{nl}(eV)\right)$ of the screened Kratzer potential for $HCl$ diatomic molecule with different values of $\varphi$.

| n | l | $\varphi = 0.2$ | $\varphi = 0.5$ | $\varphi = 0.7$ | $\varphi = 1$ |
|---|---|---|---|---|---|
| 0 | 0 | 1.254977615 | 1.286242734 | 1.307202406 | 1.338812676 |
|   |   |   |   |   |   |
| 1 | 0 | 1.296403614 | 1.391581892 | 1.456296950 | 1.555200580 |
|   | 1 | 1.279928073 | 1.375329164 | 1.440189586 | 1.539306660 |
|   |   |   |   |   |   |
| 2 | 0 | 1.338728812 | 1.500892424 | 1.612467631 | 1.784760654 |
|   | 1 | 1.322353844 | 1.484882920 | 1.596693449 | 1.769327634 |
|   | 2 | 1.289659159 | 1.452917236 | 1.565197171 | 1.738511903 |
|   |   |   |   |   |   |
| 3 | 0 | 1.381961737 | 1.614159390 | 1.775604166 | 2.027042590 |
|   | 1 | 1.365687330 | 1.598389202 | 1.760154424 | 2.012051684 |
|   | 2 | 1.333193235 | 1.566900896 | 1.729305348 | 1.982117929 |
|   | 3 | 1.284588449 | 1.519798191 | 1.683157344 | 1.937337047 |
|   |   |   |   |   |   |
| 4 | 0 | 1.426110866 | 1.731368368 | 1.945602004 | 2.281631073 |
|   | 1 | 1.409936996 | 1.715833688 | 1.930468354 | 2.267064778 |
|   | 2 | 1.377643460 | 1.684815196 | 1.900249853 | 2.237978115 |
|   | 3 | 1.329338194 | 1.638414183 | 1.855043703 | 2.194462589 |
|   | 4 | 1.265181981 | 1.576781558 | 1.794994713 | 2.136654512 |
|   |   |   |   |   |   |
| 5 | 0 | 1.471184606 | 1.852505422 | 2.122361992 | 2.548142578 |
|   | 1 | 1.455111251 | 1.837202558 | 2.107536459 | 2.533984556 |
|   | 2 | 1.423018223 | 1.806646513 | 2.077932643 | 2.505712438 |
|   | 3 | 1.375012412 | 1.760936226 | 2.033644669 | 2.463413735 |
|   | 4 | 1.311253057 | 1.700219119 | 1.974812771 | 2.407218837 |
|   | 5 | 1.231950399 | 1.624689863 | 1.901622171 | 2.337300000 |